\documentclass[twocolumn]{aastex62}

\usepackage{amsmath}
\usepackage{color}
\usepackage{amssymb}
\usepackage{float}
\addcontentsline{toc}{section}{Acknowledgement}
\usepackage{graphicx}

\newcommand{\be}{\begin{equation}}
\newcommand{\ee}{\end{equation}}
\newcommand{\bq}{\begin{eqnarray}}
\newcommand{\eq}{\end{eqnarray}}

\def\({\left(}
\def\){\right)}

\begin{document}

\title{Using the Mark Weighted Correlation Functions to Improve the Constraints on Cosmological Parameters}

\author{Yizhao Yang}, 
\affiliation{School of Physics and Astronomy, Sun Yat-Sen University, Guangzhou 510297, P.R.China}

\author{Haitao Miao}  
\affiliation{School of Physics and Astronomy, Sun Yat-Sen University, Guangzhou 510297, P.R.China}
\affiliation{Key Laboratory of Space Astronomy and Technology, National Astronomical Observatories, Chinese Academy of Sciences, Beijing 100101, China}

\author{Qinglin Ma} 
\affiliation{School of Physics and Astronomy, Sun Yat-Sen University, Guangzhou 510297, P.R.China}

\author{Miaoxin Liu}
\affiliation{School of Physics and Astronomy, Sun Yat-Sen University, Guangzhou 510297, P.R.China}

\author{Cristiano G. Sabiu}
\affiliation{Department of Astronomy, Yonsei University, 50 Yonsei-ro, Seodaemun-gu, Seoul, 03722, Korea}

\author{Jaime Forero-Romero}
\affiliation{Departamento de F{\'i}sica, Universidad de los Andes, Cra. 1 No. 18A-10 Edificio Ip, CP 111711, Bogot{\'a}, Colombia}

\author{Yuanzhu Huang}
\affiliation{School of Physics and Astronomy, Sun Yat-Sen University, Guangzhou 510297, P.R.China}

\author{Limin Lai} 
\affiliation{School of Physics and Astronomy, Sun Yat-Sen University, Guangzhou 510297, P.R.China}

\author{Qiyue Qian} 
\affiliation{School of Physics and Astronomy, Sun Yat-Sen University, Guangzhou 510297, P.R.China}

\author{Yi Zheng}
\affiliation{School of Physics and Astronomy, Sun Yat-Sen University, Guangzhou 510297, P.R.China}

\author{Xiao-Dong Li}
\affiliation{School of Physics and Astronomy, Sun Yat-Sen University, Guangzhou 510297, P.R.China}


\correspondingauthor{Xiao-Dong Li, Cristiano G. Sabiu}




%


\begin{abstract}
We used the mark weighted correlation functions (MCFs), $W(s)$, to study the large scale structure of the Universe.
We studied five types of MCFs with the weighting scheme $\rho^\alpha$, 
where $\rho$ is the local density, and $\alpha$ is taken as $-1,\ -0.5,\ 0,\ 0.5$, and 1. 
We found that different MCFs have very different amplitudes and scale-dependence.
Some of the MCFs exhibit distinctive peaks and valleys that do not exist in the standard correlation functions.
Their locations are robust against the redshifts and the background geometry, 
however it is unlikely that they can be used as ``standard rulers'' to probe the cosmic expansion history.
Nonetheless we find that these features may be used to probe parameters related with the structure formation history,
such as the values of $\sigma_8$ and the galaxy bias. 
Finally, after conducting a comprehensive analysis using the full shapes of the $W(s)$s and $W_{\Delta s}(\mu)$s,
we found that, combining different types of MCFs can significantly improve the cosmological parameter constraints.
Compared with using only the standard correlation function,
the combinations of MCFs with $\alpha=0,\ 0.5,\ 1$ and $\alpha=0,\ -1,\ -0.5,\ 0.5,\ 1$ can 
improve the constraints on $\Omega_m$ and $w$ by $\approx30\%$ and 50\%, respectively.
%
We find highly significant evidence that MCFs can improve cosmological parameter constraints.  
\end{abstract}
\keywords{ large-scale structure of Universe --- dark energy --- cosmological parameters }







\section{Introduction}\label{intro}


The discovery of cosmic acceleration \citep{riess1998observational,perlmutter1999measurements} implies
either the existence of a ``dark energy'' component in our Universe
or the breakdown of general relativity on cosmological scales. 
The theoretical explanation and observational probes of cosmic acceleration have attracted tremendous attention,
and are still far from being well understood or accurately measured \citep{weinberg1989cosmological,miao2011dark,YOO_2012,weinberg2013observational}.


On scales of a few hundred Megaparsecs (Mpc) the spatial distribution of galaxies 
forms a distinct, very complicated filamentary motif known as the `cosmic web'
\citep{1986Bardeen,1986deLapparent,
Huchra_2012,Tegmark_2004,Guzzo_2014}.
The distribution and clustering properties of galaxies in the cosmic web
encodes a huge amount of information on the expansion and structure growth history of the Universe. 
In the next decade, several large scale surveys, including 
DESI\footnote{https://desi.lbl.gov/}, EUCLID\footnote{http://sci.esa.int/euclid/}, LSST\footnote{https://www.lsst.org/},
WFIRST\footnote{https://wfirst.gsfc.nasa.gov/}, and CSST \citep{Gong_2019},
will begin operations and map out an unprecedented large volume of the Universe with extraordinary precision. 
It is of essential importance to develop powerful tools that can
comprehensively and reliably infer the cosmological parameters
from large-scale structure (LSS).

The most widely-adopted LSS analysis methods is still the
2-point correlation function (2pCF) or power spectrum measurements,
which are sensitive to the geometric and structure growth history of the Universe
\citep{kaiser1987clustering,ballinger1996measuring,Eisenstein_1998,Blake_2003,Seo_2003}.
These methods have achieved tremendous success when applied to
a series of galaxy redshift surveys such as the 2-degree Field Galaxy Redshift Survey
(2dFGRS; \cite{2df:Colless:2003wz}),
the 6-degree Field Galaxy Survey (6dFGS; \cite{beutler20116df}),
the WiggleZ survey \cite{blake2011wigglez,blake2011wigglezb},
and the Sloan Digital Sky Survey (SDSS;
\cite{york2000sloan,Eisenstein:2005su,Percival:2007yw,
anderson2012clustering,sanchez2012clustering, sanchez2013clustering, 
anderson2014clustering, samushia2014clustering,ross2015clustering,
beutler2016clustering,sanchez2016clustering,
alam2017clustering,chuang2017clustering}.
The main limitation of this method is that they are only sensitive to the gaussian part of the density field,
while both the structure formation process or some primordial conditions 
can introduce non-gaussian features in the LSS.



\begin{figure}
      \centering
      \includegraphics[width=8cm]{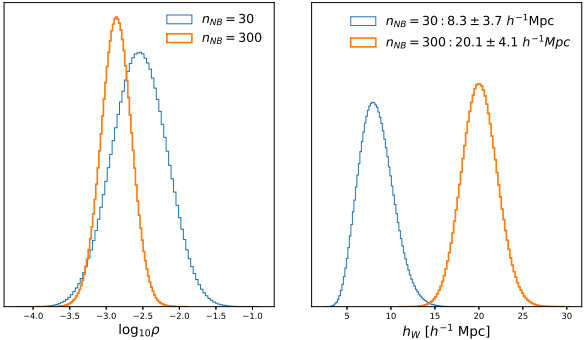}
      \caption{PDFs (probability distribution functions) of the $\log$ density and the smoothing scale $h_W$, 
      in case of using $n_{NB}$=30 and 300, measured from the BigMD $z=0.102$ halo sample.
      A masscut is applied to the sample to maintain a number density of $10^{-3}(h^{-1}{\rm Mpc})^{-3}$.
      Very roughly we find $h_{W}\varpropto n_{NB}^{0.4}$.
      }      \label{fig_hists}
\end{figure}

\begin{figure*}
      \centering
      \includegraphics[width=18cm]{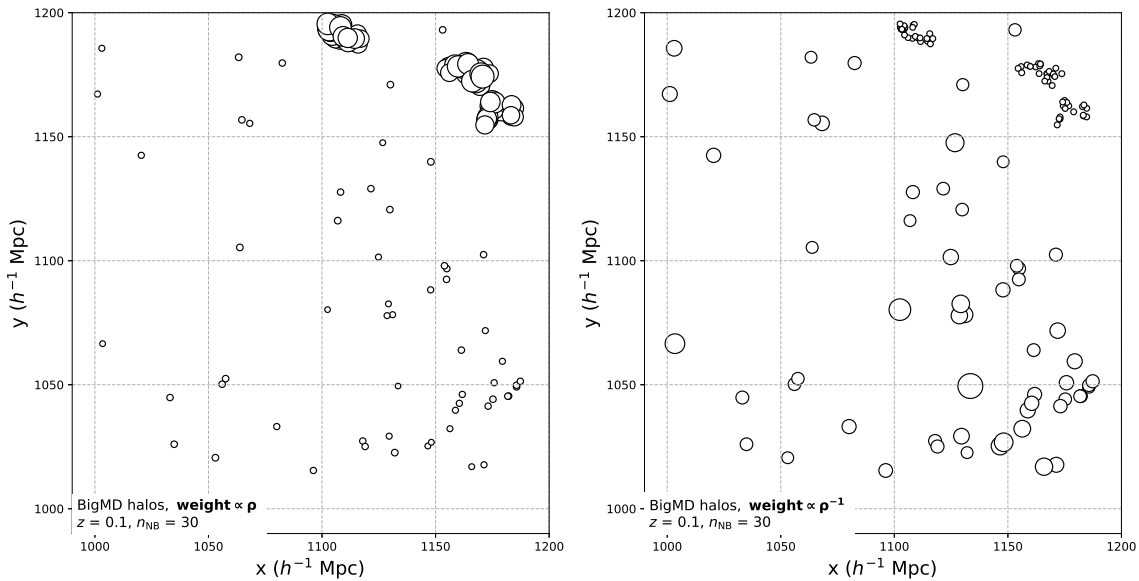}
      \caption{ 
      Halos selected from a $200\times200\times20 (h^-1\rm Mpc)^3$ slice in the BigMD sample.
      Their weights, as represented by the circle size, 
      are determined by $\rho^{\alpha}$, where $\alpha=1,\ -1$. 
      The $\alpha=1$ scheme puts significantly more emphasis on
      the objects in dense environment, 
      while the other scheme does the opposite. 
      }      \label{fig_scatter}
\end{figure*}

Ongoing research seeks to go beyond the 2-point statistics
includes the methods such as 3-point statistics \citep{Sabiu_2016,Slepian_2017},
4-point statistics \citep{Sabiu_2019},
cosmic voids \citep{ryden1995measuring,lavaux2012precision},
deep learning \citep{Ravanbakhsh17,Mathuriya18},
and so on.
While many of them have proved useful,
here we investigate another statistical tool, namely the {\em mark weighted correlation function} 
\citep[MCF; ][]{Beisbart2000,Beisbart2002,Gottl2002,Sheth:2004vb,Sheth:2005aj,Skibba2006,White_2009,White_2016,Satpathy:2019nvo,2020arXiv200111024M,PMS2020}
which is simpler and computationally easier compared than the statistics mentioned above.

By weighting each galaxy using a ``mark'' that depends on its local 
density, the MCFs provide density-dependent clustering information from
the sample which is useful for data mining.
The weights can be set to be proportional to the positive or negative
power of the density, to allow the statistics to place more emphasis
on dense or undense regions,
where the clustered structures and the redshift space distortions (RSDs) are physically very different.
It is expectable that in this manner we can obtain more information from the data
compared with using the traditional 2pcf,
which  equally treat all galaxy pairs regardless of 
the difference in their physical properties and environments.


This paper is arranged as follows. 
Section \ref{data} outlines the parameters of the datasets we use. 
Section \ref{method} represents the methods used for the principle and operation of marked correlation functions. 
Section \ref{basic} explains the clustering statistics of different number density. 
Section \ref{discussion} discusses more details by various parameters to test whether the standard ruler persists. 
In section \ref{conclusion} we present our general conclusions.


\section{Data}
\label{data} 


The analysis in this work relies on the large N-body simulation: BigMultiDark\footnote{webpage: https://www.cosmosim.org } (BigMD),
and also a series of fast simulations generated using COLA (COmoving Lagrangian Acceleration).

The Multiverse simulations are a set of cosmological N-body simulations designed to 
study how variations in cosmological parameters affect the clustering and evolution of cosmic structures. 
Among them, the BigMD simulation is produced using $3\,840^3$ particles in a volume of $(2.5h^{-1}\rm Gpc)^3$,
assuming a $\Lambda$CDM cosmology with
 $\Omega_m = 0.307115$, $\Omega_b = 0.048206$, $\sigma_8 = 0.8288$, $n_s = 0.9611$, and $H_0 = 67.77\ {\rm km}\ s^{-1} {\rm Mpc}^{-1}$ \citep{BD}.
The initial conditions, based on primordial Gaussian fluctuations, 
are generated via the Zel'dovich approximation at $z_{\rm init} = 100$. 
Its large volume and huge number of particles make this an ideal simulation for 
the purposes of this work.



For purpose of estimating measurement covariance matrices, 
we also use 150 simulations with $600^3$ particles and a boxsize $(512 h^{-1}\rm Mpc)^3$, in the BigMD cosmology,
generated using the COLA \citep{Tassev13} algorithm.
Second order Lagrangian perturbation Theory (2LPT) is a computationally efficient and accurate method for describing the gravitational dynamics on large scales. 
COLA combines 2LPT, for time integration for large scale dynamical evolution, 
with a full-blown N-body Particle-Mesh (PM) algorithm to calculate the small scale dynamics. 
Compared with the other fast simulation algorithms in the market, 
COLA performs better in simulating the structures on non-linear scales \citep{Chuang14}.



Finally, to check the dependence on cosmologies, 
we run five sets of COLA simulations 
using $\Lambda$CDM models of 
$(\Omega_m, 10^9 A_s, \sigma_8)$ = (0.2, 2.1, 0.5557), (0.31, 2, 0.7965),
(0.31, 2.1, 0.8161), (0.31, 2.29, 0.8523) and (0.46, 2.1, 1.0576) 
\footnote{Notice that these three parameters are actually not independent; 
$\sigma_8$ is usually considered as a derived parameter crucially dependent on $\Omega_m$ and $A_s$.},
respectively. 
The other parameters are taken as  $\Omega_b = 0.048206$, $n_s = 0.9611$ and $H_0 = 67.77\ {\rm km}\ s^{-1} {\rm Mpc}^{-1}$,
the same to their used values in the BigMD simulation.
Each simulation was run using $1024^3$ particles in a $(1024 h^{-1}\rm Mpc)^3$ box.


We identify gravitationally bound structures in each of the BigMD and COLA DM simulations using the \textsc{ROCKSTAR} halo finder \citep{ROCKSTAR}.
\textsc{ROCKSTAR} is a halo finder based on adaptive  hierarchical refinement  of  friends-of-friends  groups  
in six phase-space dimensions  and  one time dimension, 
allowing for robust tracking of substructure.
Both halos and subhalos are included in the analysis.
To ensure the comparability, we maintain a halo number density $\bar n=$0.001 $(h^{-1}\rm Mpc)^{-3}$ in all simulations. 

\section{Methodology}\label{method}


The MCF is a simple extension of the standard configuration space 2pCF 
by assigning a mark to each object.
Following \citep{White:2016yhs}, we use the local density as the mark,
and weight each halo by
\begin{equation}
 {\rm weight}=\rho_{n_{\rm NB}}^\alpha,
\end{equation}
which is a simpler expression than that proposed in \cite{White:2016yhs}.
Here $\rho_{n_{\rm NB}}$ is the density
estimated 
using its $n_{\rm NB}$ nearest neighbours,
\begin{equation}
 \rho_{n_{\rm NB}}({\bf r}) = \sum_{i=1}^{n_{\rm NB}}  W_k({\bf r-r_i},h_W),
\end{equation}
where $\rho_{n_{\rm NB}}(\bf r)$ is the number density at position $r$,
and $W_k$ is the smoothing kernel,
for which we choose the 3rd order B-spline functions
having non-zero value within a sphere of radius $2h_W$ $h^{-1}$ Mpc
\citep{Gingold1977, Lucy1977}.
We adopt an adjustable radius of the smoothing kernel 
to ensure that the kernel always includes $n_{\rm NB}$ nearest neighbour halos within $2h_W$.

The value of $n_{\rm NB}$ determines the smoothing scale that we applied 
to the sample.
Figure \ref{fig_hists} shows the PDF (probability distribution function) of the $\log$ density and 
$h_W$ in the constructed fields when using $n_{\rm NB}$=30 and 300, respectively.
Since the values of $h_W$ depend on the local density, they are not a constant number under a given $n_{\rm NB}$.
Here we find the central and 1$\sigma$ width of $h_W$ is $8.3\pm 3.7$, $20.1\pm 4.7$ $h^{-1}$Mpc
if using $n_{\rm NB}$=30, 300, respectively.
Very roughly, the central value scales as $\varpropto (n_{\rm NB})^{0.4}$. 
A larger $n_{\rm NB}$ decreases both the mean and the variance of $\log_{10}\rho$.

In the MCF, the objects in the high and low dense regions are assigned different weights. 
Figure \ref{fig_scatter} shows the weights 
of some halos distributed in a $200\times200\times20 (h^{-1}\rm Mpc)^3$ slice, 
selected from the $z=0.102$ BigMD snapshot. 
While $\alpha=1$ assign significantly larger weights to
the objects in dense environment, 
the $\alpha=-1$ strategy does the opposite.
From the dense to un-dense regions,
the clustering patterns and redshift space distortions vary dramatically,
so we expect very different results for MCFs when using the two weighting strategies.


Apart from the weight that is assigned to each halo,
the computational procedure to measure the MCF
is exactly the same as that to measurement the standard 2pCFs.
We use the most commonly adopted Landy-Szalay estimator 
\begin{equation}\label{eq:Wsmu}
 W(s,\mu) = \frac{WW-2WR+RR}{RR},
\end{equation}
where $WW$ is the weighted number of galaxy-galaxy pairs,
$WR$ denote the galaxy-random pairs,
and $RR$ denote the number of random-random pairs. 
They are separated by a distance defined by 
$s\pm \Delta s$ and $\mu\pm \Delta \mu$,
where $s$ is the distance between the pair and
$\mu=\cos(\theta)$, with $\theta$ being the angle 
between the line joining the pair and the line of sight (LOS) direction
\footnote{Here we use $s$ instead of $r$ because the statistics is usually performed 
using the redshift space positions,
due to the RSDs they are related with each other via $s = r+v/(aH)$.}.
For the random samples, we always use 10x more particles than the data samples,
and fix the weights of all particles to be 1.

Compared to the tradition CF which is defined as $\xi(\bf r)=\left<\delta({\bf x})  \delta({\bf x+r})\right>$,
the MCF takes the form of
\begin{eqnarray}\label{eq:W_delta_rho}
 W({\bf r}) &=& 
 \left<\delta({\bf x}) \rho_{n_{NB}}({\bf x})^{\alpha} \delta({\bf x+r}) \rho_{n_{NB}}({\bf x+r})^{\alpha}\right> .
\end{eqnarray}
Notice the difference between $\delta$ and $\rho_{n_{\rm NB}}$.
The latter one is the smoothed density field,
while the former is the contrast of the {\it point-like} density $\rho$.
effectively, $\rho$ is the special case of $\rho_{n_{\rm NB}}$ with $n_{\rm NB}$=1.

\section{A glance at the weighted CF}
\label{basic}


In what follows we present the MCFs measured from the BigMD halos, 
distributed in the redshift range of $0<z<1.45$.
To guarantee the comparability of objects at different redshifts,
and also to maintain a uniform smoothing scale, 
from each sample we select a number of most massive halos to build up 
a subsample having a constant number density 
$\bar n=10^{-3} (h^{-1} \rm Mpc)^{-3}$ for all samples.

\subsection{$W(s)$ Measurements}\label{sec:ws}

\begin{figure*}
      \centering
      \includegraphics[width=16cm]{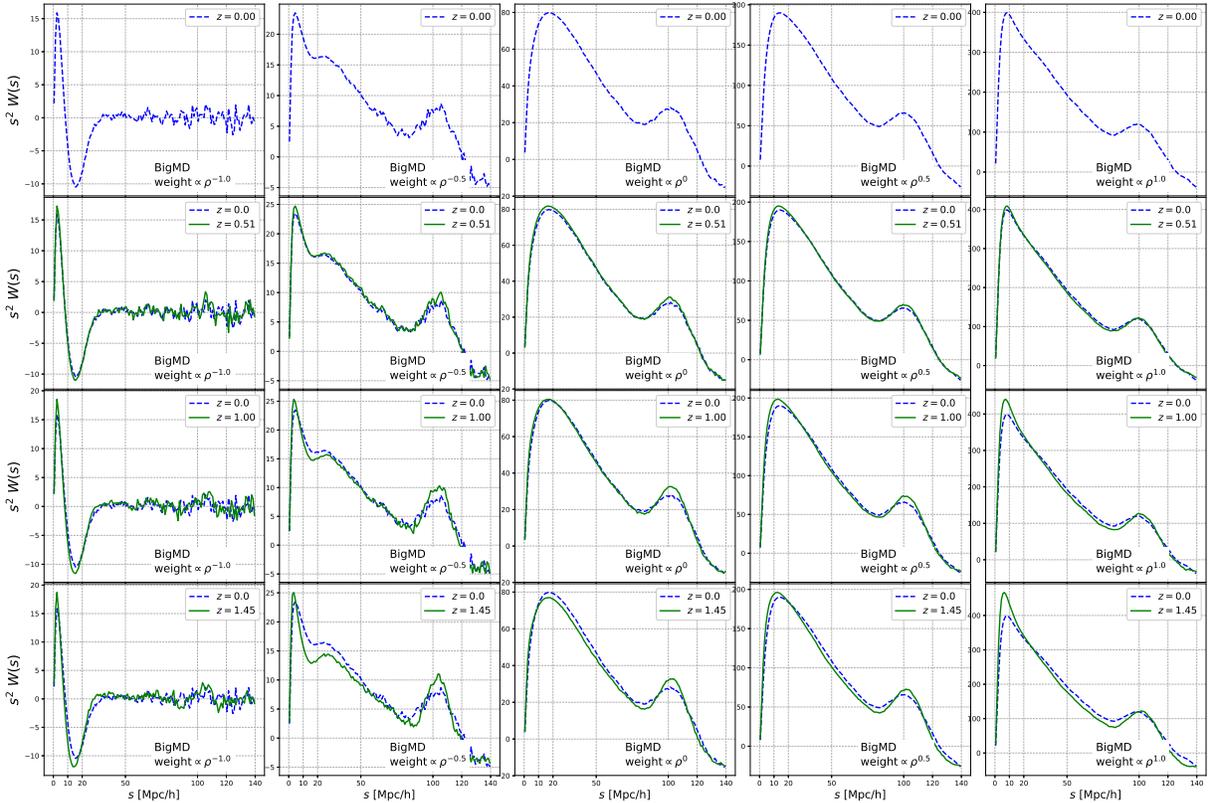}
      \caption{
      MCFs of BigMD halos measured using weight$\propto\rho_{30}^{\alpha}$, 
      where we use $\alpha = -1, -0.5, 0, 0.5, 1$, respectively.
      From top to bottom, measurements 
      at redshifts of 0, 0.51, 1.0 and 1.45 are presented. 
      Both the shapes and amplitudes of the MCFs are 
      sensitive to the choice of $\alpha$.
      See context of Sec. \ref{sec:ws} for more discussions.
      }\label{fig_xi_s}
\end{figure*}

Figure \ref{fig_xi_s} shows the MCFs as functions of clustering scale, i.e. the monopole $W(s)$
\footnote{Here we do not study the higher order multipoles, 
since the $\mu$-dependence is studied in the next Section using another statistical quantity.}.
They are computed by ignoring the $\mu$-dependence in Equation \ref{eq:Wsmu}
when counting the weighted number of pairs.
We shows the results using $\alpha=-1,\ -0.5,\ 0,\ 0.5$, and $1$,
at the redshifts of 0, 0.51, 1.0, and 1.45, respectively.
In all plots, we use $n_{NB}=30$.

\begin{figure*}
      \centering
      \includegraphics[width=16cm]{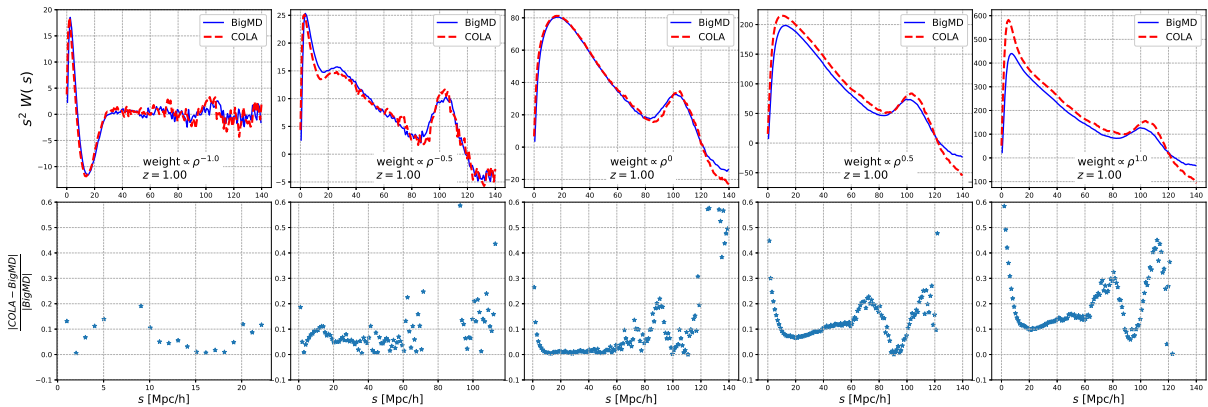}
      \caption{Comparing the MCFs measured from the BigMD  and COLA samples. 
      COLA achieves $\lesssim10\%$ accuracy except the 
      $s\lesssim20$ $h^{-1}$Mpc regime in the $\alpha=0.5$ and 1 cases.
      The COLA measurements are used for the estimation of covariance matrix.
      }\label{fig_bigmd_vs_cola}
\end{figure*}


A significant dependence on the weighting scheme is detected 
when comparing the MCFs using different $\alpha$.
A larger $\alpha$ assigns more weights to the dense, clustered region,
thus results in stronger correlation (higher magnitude). 
The clustering patterns in dense and undense regions are different from each other,
so the shape of MCFs is also sensitive to $\alpha$.

As shown in the Figure, when using $\alpha=-1,\ -0.5,\ 0,\ 0.5$ and $1$, 
$s^2 W(z=0)$ peaks at $s\approx2,\ 4,\ 17,\ 14$ and $18$ $h^{-1}\ $Mpc,
with amplitudes of 16, 23, 80, 180, 400 $(h^{-1}\rm Mpc)^2$, respectively.
The $\alpha=1$ result has a peak magnitude 5 times stronger than the $\alpha=0$ case, 
while the latter is again 5 times stronger if compared with the $\alpha=-1$ case;
if comparing the clustering amplitude on the BAO scale, 
then the $\alpha=1$ case is 2/4/15/100 times stronger than 
$\alpha=0.5/0/-0.5/-1$ cases, respectively.
The statistical error also increases with the decreasing of $\alpha$.
For the $\alpha=-1$, $z=0$ case, the BAO peak is not very detectable, possibly due to the large noise therein.

By enforcing $\bar n=10^{-3}$ at all redshifts 
both the clustering amplitude and the shape remain similar at all redshifts.
Compared with low redshift result we find the BAO peak at higher redshift is more prominent,
because there the smearing effect from the peculiar velocity and the non-linear structure formation is less significant.

The shape of the MCF is changing persistently when we tune the value of $\alpha$.
Several distinctive features, 
including a sharp peak (around 5-10 $h^{-1}$ Mpc) 
in the $\alpha=1,\ -0.5,\ -1$ results,
are a valley (around $15 h^{-1}$ Mpc) in the $\alpha=-0.5,\ -1$ results,
are detected.
We will discuss their origins, implications and usabilities in the latter sections.

Finally, a quick check presented in Figure  \ref{fig_bigmd_vs_cola} shows that 
for most cases COLA achieves $\lesssim10\%$ accuracy in predicting the MCFs within 
the clustering range considered here \cite{Ma_2020}.
Relative large discrepancy is detected 
at the $s\lesssim20$ $h^{-1}$Mpc
regime in the $\alpha=0.5$ and 1 cases.
This consistency may be resolved by measures such as increasing the time steps or enhancing the resolution of the simulations,
but we will not study it in details.

\begin{figure*}
      \centering
      \includegraphics[width=16cm]{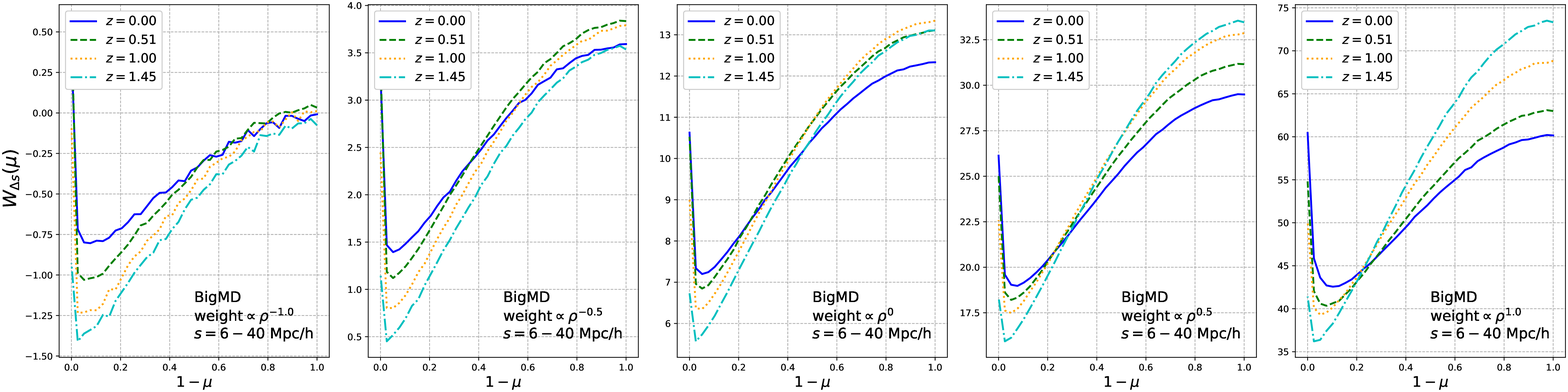}
      \includegraphics[width=16cm]{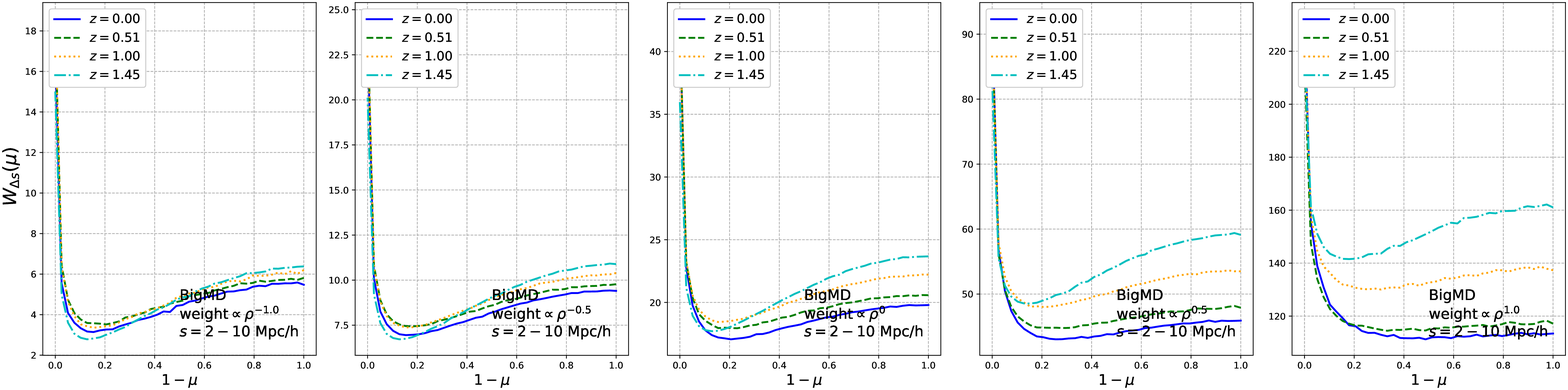}
      \includegraphics[width=16cm]{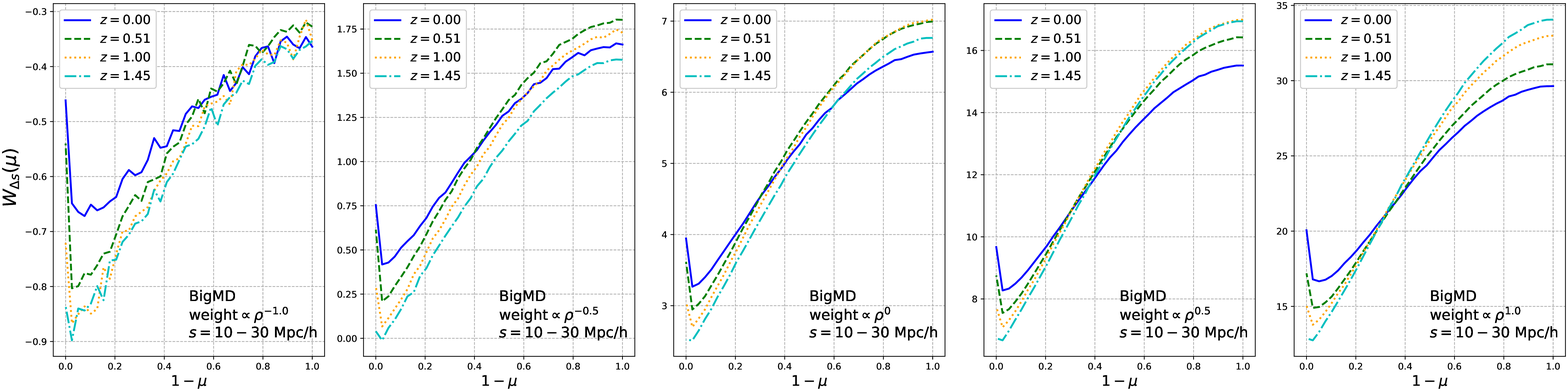}
      \caption{
      Anisotropic clustering in the MCFs.
      Shown in the figures are $W_{\Delta s}\equiv\int_{s_{\rm min}}^{s_{\rm max}} W(s,\mu)ds$,
      where the integration ranges are (6,40), (2,10) and (10,30) $h^{-1}$ Mpc, respectively. 
      In all plots, there is a sharp peak near $1-\mu=0.1$ as produced by the FOG effect,
      and a slope in the range of $1-\mu\gtrsim0.1$ as created by the Kaiser effect.
      The shape and the amplitude depends on the value of $\alpha$.
      } \label{fig_ximu}
\end{figure*}

\subsection{$W_{\Delta s}(\mu)$ Measurements}\label{sec:wmu}


The RSDs in high and low density regions are quite different.
So we expect different anisotropic clustering features in the different MCFs.
In what follows, we study $\mu$-dependence of the MCFs.
By integrating $W(s,\mu)$ along the $s$ direction, we define 
\begin{equation}
 W_{\Delta s}(\mu)\equiv \int_{s_{\rm min}}^{s_{\rm max}} W(s,\mu) ds,
\end{equation}
as well as its normalized version
\begin{equation}
 \hat W_{\Delta s}(\mu)\equiv \frac{W_{\Delta s}(\mu)}{\int_{0}^{\mu_{\rm max}}W_{\Delta s}(\mu)\ d\mu}.
\end{equation}
These two quantities describe the difference in the clustering strength in different directions {\em w.r.t} the LOS.
They have been used to quantify the RSDs and the AP distortions in the tomographic Alcock-Pacyznski (AP) method \cite{LI14,LI15,LI16,LI18,LI19,Park:2019mvn,Zhang2019}.

Figure \ref{fig_ximu} shows the measured $W_{\Delta s}(\mu)$ at redshifts of $z=$ 0, 0.51, 1 1.45,
using $\alpha$=-1, -0.5, 0, 0.5, 1 and an integral range 
$s\in$ (6,40), (2,10), (10,30) $h^{-1}$ Mpc, respectively.
In all curves, we see a sharp peak near $1-\mu=0.1$,
which is produced by the small-scale, non-linear finger-of-god (FOG) effect \citep{Jackson_1972};
Also, we see a slope in the range of $1-\mu\gtrsim0.1$,
as a consequence of the Kaiser effect \citep{kaiser1987clustering}.

The amplitude of $W_{\Delta s}(\mu)$ is enhanced
if we tune down $s_{\rm min}$ 
 and include more small-scale clustering into the integration.
In doing this, we also enhance the leftmost peak 
since FOG is stronger on smaller clustering scales. 

Similar to what we found with $W(s)$,
the  $W_{\Delta s}(\mu)$ has a larger amplitude and smaller statistical noise when using a larger value of $\alpha$.
On the other hand, 
we do not detect any ``violent'' changes in the shape of $W_{\Delta s}(\mu)$ when tuning the value of $\alpha$.
However, this does not necessarily mean that the information encoded in these different $W_{\Delta s}(\mu)$ are all the same.
We will revisit this issue in the later.


\subsection{Distinctive Features in $W(s)$}\label{sec:distinctive_features}


In the $W(s)$ curves there are several distinctive peaks and valleys
which do not exist in the standard 2pCFs.
In what follows, we briefly discuss their possible origins.

\begin{figure}
      \centering
      \includegraphics[width=8cm]{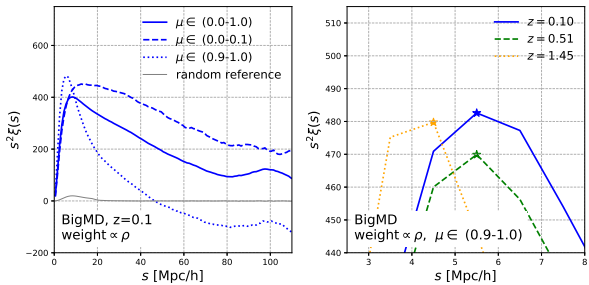}
      \caption{
      In the $\alpha$=1 MCF, we find a sharp peak around $s=5\ h^{-1}$ Mpc.
      Comparison between the measurements on different directions implies that this peak has something to do with the FOG effect (left panel).
      Its location and position is insensitive to the redshift (right panel).
      } \label{fig_rho1}
\end{figure}

\begin{figure}
      \centering
      \includegraphics[width=8cm]{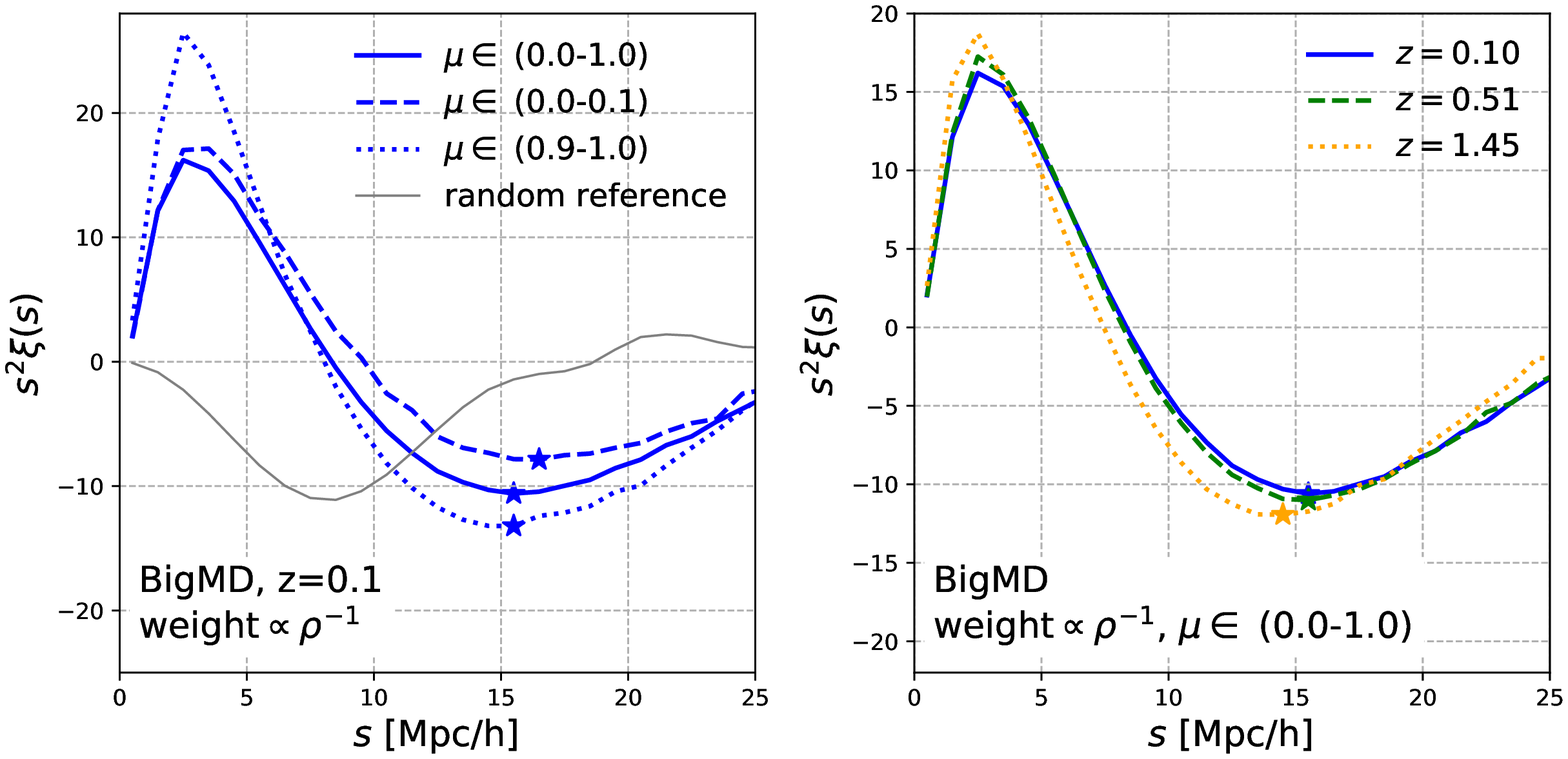}
      \caption{
      The $\alpha$=-1 MCF has an unusual ``S''-like shape.
      It bears a valley-like feature on scales of $\sim$15 ${h^{-1}}$ Mpc, 
      arising from the difference between the point-like field $\rho$ and the smoothed field $\rho_{30}$.
      Its shape and the strength of remains robust against redshift.} \label{fig_rho-1}
\end{figure}

\subsubsection{Sharp peak}\label{section_sharp_peak}

In many plots of $s^{2}W(s)$ there exist a sharp peak located around $5-10\ h^{-1}$ Mpc (see Figure \ref{fig_xi_s}).
This means that on that scale there exist a large 
number of clustering pairs.


In all plots we use the weight $\rho_{30}$, 
whose smoothing scale is $\approx 8$ $h^{-1}$ Mpc.
That smoothing produces a correlation on that scale,
so it is not surprising to see a peak on the corresponding scale.
However, comparing with a random sample smoothed in the same way
shows that the the amplitude of the peak also 
heavily depends on the intrinsic clustering property of the sample. 


A comparison between the measurements on different directions implies that
this sharp clustering peak has something to do with the FOG effect,
which produces $\sim5-10\ h^{-1}$ Mpc ``spikes'' like structure along the LOS 
As shown in the left panel of Figure \ref{fig_rho1},
the peak along the LOS direction is far more prominent than what found in the transverse direction.
The other panel of the Figure shows that
the heights and locations of the peak is rather insensitive to the redshift.





\subsubsection{Distinctive Valley}

Besides the peak we also detect a valley located at $s\approx15\ h^{-1}$ Mpc
in the $\alpha=-1,\ -0.5$ cases.
In particular, the $\alpha=-1$ case possesses both the peak and the valley,
so has an unusual ``S''-like shape.

Figure \ref{fig_rho-1} shows the valleys in the $\alpha=-1$ MCFs.
Equation \ref{eq:W_delta_rho} means that 
the features in the MCF should be highly related with 
the difference between the local density $\rho$ 
and its smoothed counterpart $\rho_{30}$.
In the redshift range of $\sim$0-1.5, 
the location and the strength of this valley-like feature remains rather robust.


Contrary to the situation of the peak,
we find that the valley looks rather similar in both the LOS and transverse directions,
leading us to believe that it has little or nothing to do with RSD effects.

\section{Implications for Cosmological Analysis}\label{discussion}


In this section we discuss the implications of the MCFs to the cosmological analysis.
In the first part, we report our attempt to utilise the peak and the valley 
as standard rulers.
In the second part, we adopt a more comprehensive approach 
by using the full shape of the MCFs.

\subsection{Usability of the distinctive features as standard rulers }\label{subsec:features}

The distinctive peaks and valleys as discovered in the $\alpha \neq 0$ MCFs 
are not found in the standard 2pCF.
A remarkable feature is that, the locations of these peaks and valleys 
are rather robust against the redshift.
This inspires us to consider using them as ``standard rulers'' to probe the expansion history.




In galaxy surveys, the angular positions and redshifts of each galaxy is converted to 3D positions 
using the redshift-distance relation $r(z)$ adopted in an assumed cosmology.
So wrongly adopted cosmology parameters lead to the following
distortions of length in the directions parallel and perpendicular to the LOS,
\begin{equation}\label{eq:alpha}
	\begin{aligned}
	&\alpha_{\parallel}(z) = \frac{H_{\rm true}(z)}{H_{\rm wrong}(z)}, \\
	&\alpha_{\perp}(z) = \frac{D_{A,\rm wrong}(z)}{D_{A,\rm true}(z)},
	\end{aligned}
\end{equation}
where ``true'' and ``wrong'' denote the values of quantities in the true and incorrectly assumed cosmologies, respectively.
This leads to two effects in the wrong cosmology,
\begin{itemize}
 \item The changes in the size of structures, known as the ``volume effect''.
 This changes the BAO peak location, 
 shifts the clustering patterns \citep{Li2017},
 and changes the sizes of structures in the density field \citep{Park2010}. 
 \item Changes in the shape of structures, known as the Alcock-Paczynsk (AP) distortion \citep{AP1979,ballinger1996measuring}.
  For an incomplete list of the methods based on this effect and their applications to the data, see
  \cite{Ryden1995,Matsubara1996,Outram2004,Marinoni2010,Blake2011,Lavaux2012,Alam2017,Qingqing2016,LI14,LI16,Ramanah2019}.
\end{itemize}
In what follows, we mainly test the feasibility of using the distinctive features 
to probe the ``volume effect''. 
To mimick the effect,
we take Equation \ref{eq:alpha} to convert the sample into backgrounds 
of two wrong cosmologies, 
\begin{equation}
 (\Omega_m,\ w)_{\rm wrong} = (0.1,\ -1), \ \ (0.3071,\ -1.5).
\end{equation}
Notice that, in doing this we just ``re-observe'' the simulation using the $r(z)$s of the new cosmologies,
without running new simulations.
That is exactly what one is doing when conducting the BAO or AP analysis on the observational data.

The comparability of samples requires them having the same smoothing scale. 
Thus, when using a wrong background, we change the lower halo mass cut to maintain a constant number density $\bar n=1\times10^{-3} (h^{-1}\rm Mpc)^{-3}$.

In the following subsections, we test the 
feasibility of using the $\alpha=-1$ and $\alpha=1$ MCFs, respectively.






\subsubsection{The background}

\begin{figure}
      \centering
      \includegraphics[width=9cm]{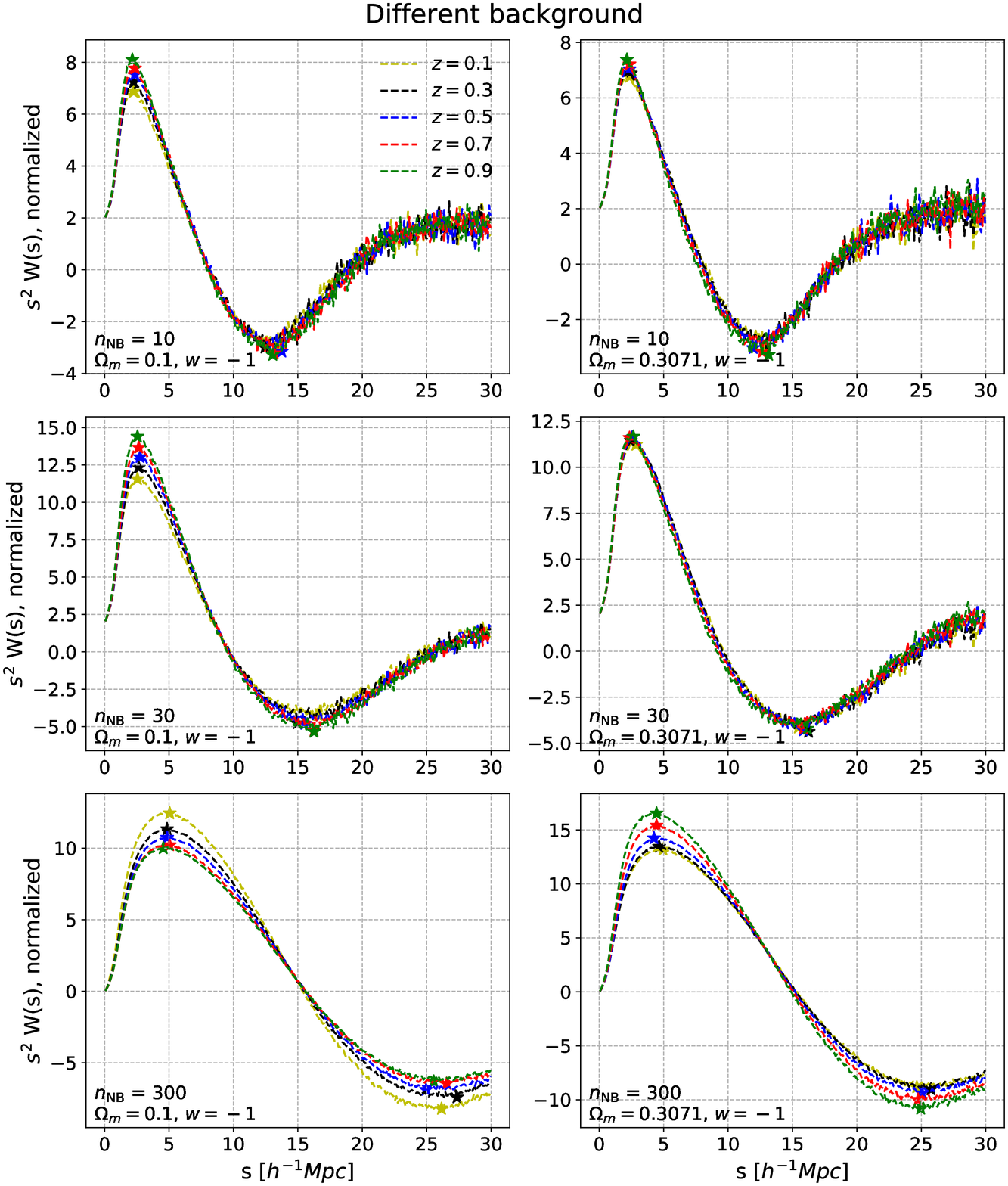}
      \caption{Comparing the $\alpha=-1$ MCFs in backgrounds of 
      the fiducial cosmology and an extremely wrong cosmology. 
      Regardless of the dramatic difference in the background geometry,
      there is little difference in the locations of 
      the peaks or the valleys.
      So it is unlikely to use these features to probe the geometry of the Universe.
      }\label{fig_peak_wrong_cosmos}
\end{figure}

\begin{figure}
      \centering
      \includegraphics[width=9cm]{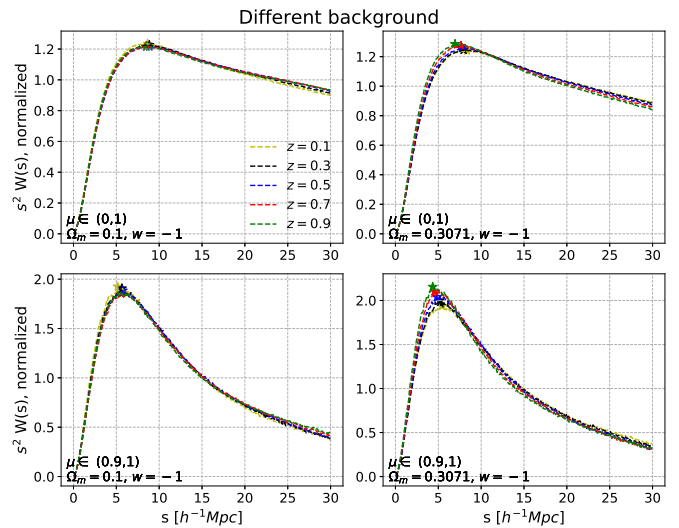}
      \caption{
      Comparing the $\alpha=1$ MCFs in different backgrounds.
      The conclusion is similar to what we found in Figure \ref{fig_peak_wrong_cosmos}.
      } \label{fig_peak_wrong_cosmos_rho1}
\end{figure}

When adopting an incorrect expansion history for the background, we expect 
a scale-shift in the shape of the CF.
Because of the nonuniform scaling at different redshifts,
we expect a redshift-evolution of the CFs,
determined by $\left(\alpha_{\perp}(z)^2\alpha_{\parallel}(z)\right)^{1/3}$ (see \cite{Li2017}).


However, regardless of the strong volume effect in the two extremely incorrect cosmologies considered here,
we do not detect any significant change in the scale of the MCFs.
Figure \ref{fig_peak_wrong_cosmos} shows that,
in the wrong cosmological backgrounds,
the locations of the peaks or valleys remain the same to their fiducial values.
The conclusion is unchanged when we try using $n_{\rm NB}=10$, 30 and 300.

Considering the valleys of the $n_{\rm NB}$=30 measurements as an example.
While in the fiducial cosmology the valley locates at $s=15-16$ $h^{-1}\rm Mpc$,
in the $\Omega_m=0.1$ wrong cosmology it still shows up at $s\approx16$  $h^{-1}\rm Mpc$. 
For comparison, in this cosmology the comoving length is artificially rescaled by
a rate of $\gtrsim$20\% at $z\gtrsim0.6$,
so we expect the valley appears near 18-19 $h^{-1}$ Mpc.

While being insensitive to the background,
the location of the peaks or valleys are rather sensitive to the choice of the smoothing scale.
When changing $n_{\rm NB}$ from 30 to 10/300, 
the location of the valley is shifted to 
26/13 $h^{-1}\rm Mpc$, respectively.

In the $\alpha=1$ MCFs, 
again, we find that the locations of the peaks are,
rather insensitive to the background change (see Figure \ref{fig_peak_wrong_cosmos_rho1}).
Moreover, it appears robust against changes in redshift
in the wrong cosmology we chose.
This means that it is impossible to make use of their redshift evolution
as a signal to identity the wrong cosmologies 
\footnote{Not only the locations of the peaks/valleys are insensitive to the background change, 
we find their heights also being rather insensitive to the background.
The reason is that, by maintaining a same number density in all backgrounds,
we are selecting objects with different bias;
the change in the bias counteracts the effect of the background alteration on the clustering strength.
}.

Here we point out that, actually, this FOG related pattern has been detected in other statistics.
\cite{Fang2019} reported a detection of a peak around $\sim3h^{-1}{\rm Mpc}$ in the $\beta$-skeleton statistics.
In Appendix A, 
we report that the peak in that statistics can not be used to conduct cosmological analysis, either.

\subsubsection{Dependence on bias and $\sigma_8$}\label{sec_five_cola_sim}

\begin{figure}
      \centering
      \includegraphics[width=9cm]{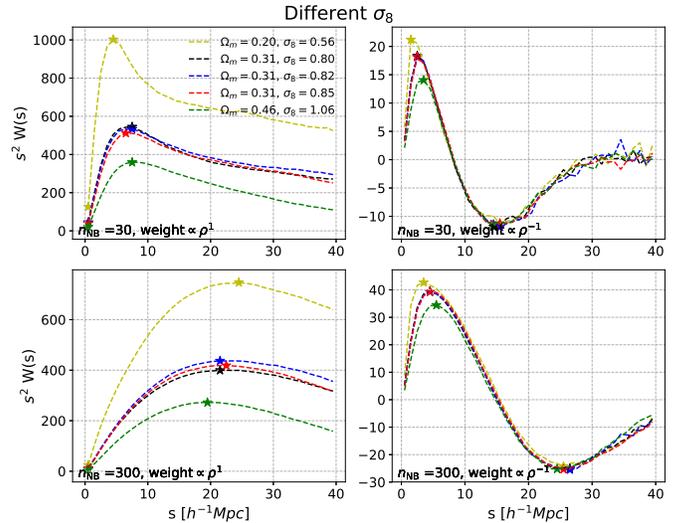}
      \caption{
      MCFs of five COLA simulations, with different values of $\sigma_8$.
      The peak is shifted towards to larger scales when 
      using a larger $\sigma_8$.
      }  \label{fig_diffsim_xis}
\end{figure}

While being rather insensitive to the background,
these features do have some dependence on the bias and $\sigma_8$.
Figure \ref{fig_diffsim_xis} shows the MCFs
measured in five sets of COLA simulations,
with the $\Lambda$CDM parameters of 
$(\Omega_m, 10^9 A_s, \sigma_8)$ = (0.2, 2.1, 0.5557), (0.31, 2, 0.7965),
(0.31, 2.1, 0.8161), (0.31, 2.29, 0.8523) and (0.46, 2.1, 1.0576),
respectively.
Clearly, when adopting a smaller $\sigma_8$,
the locations of the peaks shift towards small scales. 

Basically, a smaller $\sigma_8$ leads to a smaller peak scale,
except that the $\sigma_8=0.8/0.82/0.85$ curves in the $\alpha=1$,
$n_{\rm NB}=300$ case do not precisely obey this order.
Possibly, becase $n_{\rm NB}=300$ corresponds to a smoothing scale
much larger than 8$h^{-1}$Mpc, here $\sigma_8$ can not precisely
describe what is happening. 
Although the basic trend is still correct,
some complexities arise if we carefully investigate the details.

\begin{figure}
      \centering
      \includegraphics[width=9cm]{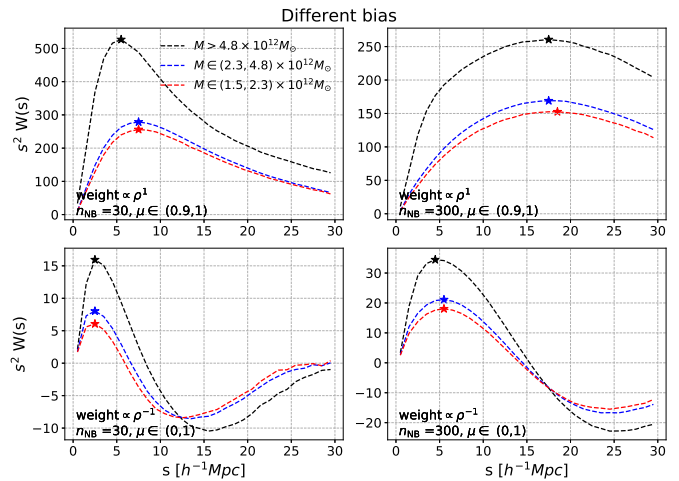}
      \caption{
      MCFs of samples with different halo bias.
      The valleys are more affected compared with the peaks.      
      }   \label{fig_diffbias}
\end{figure}

Meanwhile, we also find they have some dependence on the halo bias.
Figure \ref{fig_diffbias} shows the 
MCFs of three subsamples of BigMD $z=0$ halos,
distributed in different mass range
(we keep $\bar n=10^{-3}\ (h^{-1}\rm Mpc)^{-3}$ in all subsamples).
The valleys are more affected compared with the peaks.

In summary, our analysis shows that these peaks and valleys can not be used as ``standard rulers'' 
to probe the geometry of the Universe.
But when changing the parameters related with the structure formation, 
we do observe shifts in the peaks or valleys.
So these features maybe useful for the probing of those parameters related with the structure formation,
e.g. the values of $\sigma_8$ and the halo/galaxy bias.

Apart from $\hat W(s)$s, the $\hat W_{\Delta s}(\mu)$s are also sensitive to $\sigma_8$.
We investigate this sensitivity further in Appendix B.

\subsection{Using the full shape of the MCFs}

\begin{figure*}
      \centering
      \includegraphics[width=8cm]{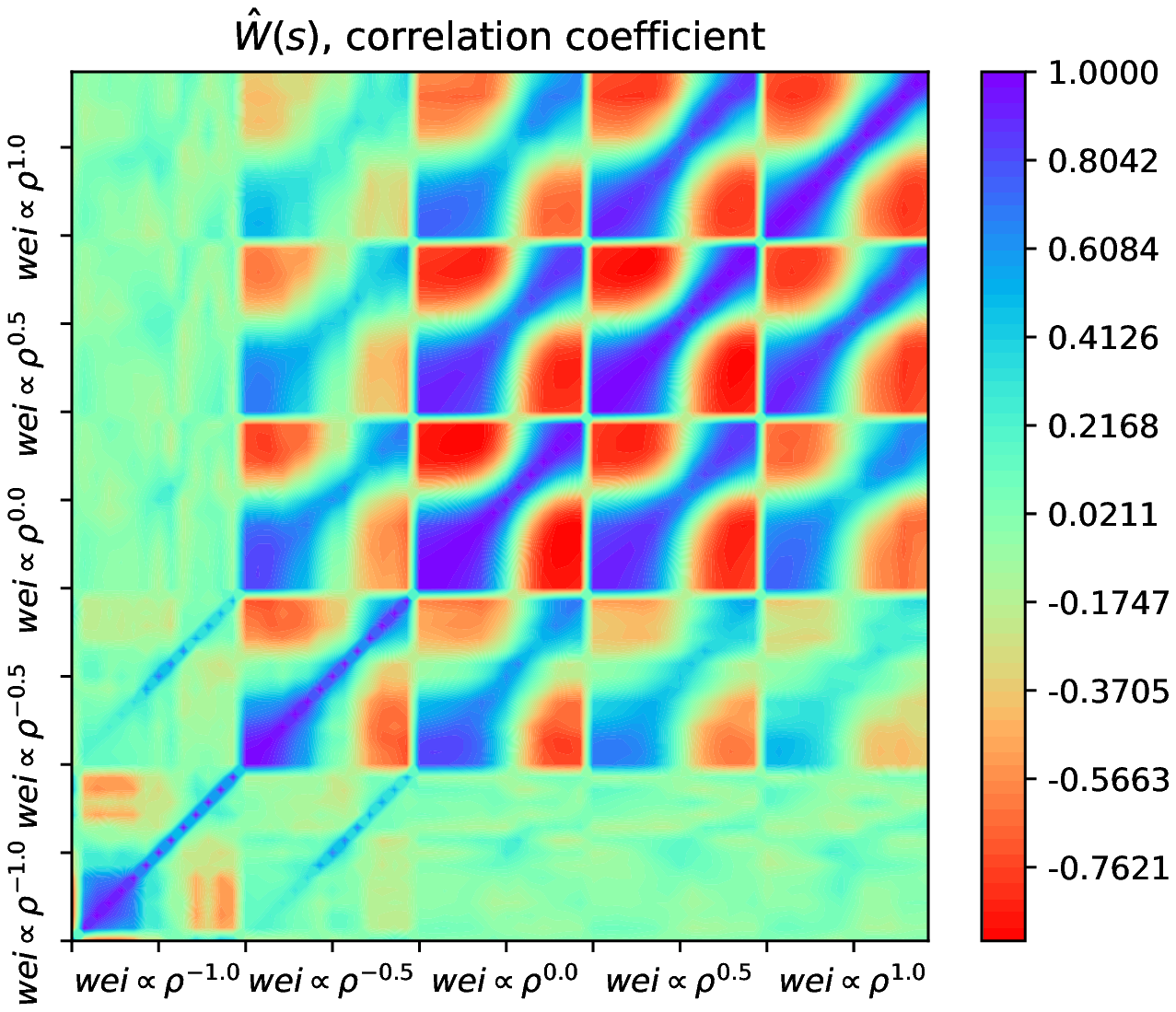}
      \includegraphics[width=8cm]{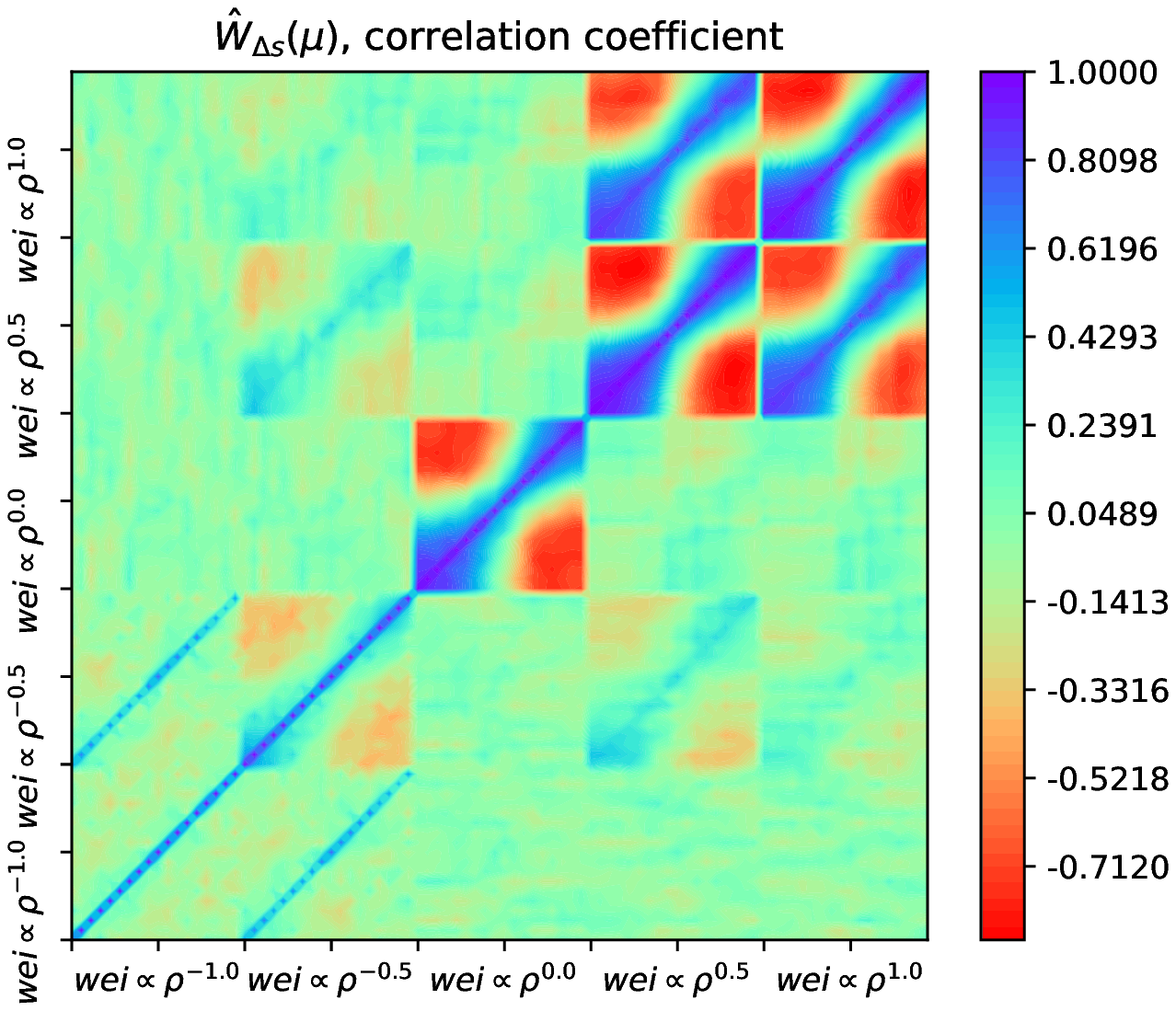}
      \caption{The correlation coefficients of$\hat W(s)$ and $\hat W_{\Delta s}(\mu)$,
      computed using the 150 sets of COLA simulations.
      The $\alpha=-1$ MCF has weakest correlation with the others.
      }      \label{fig_cov}
\end{figure*}

In what follows, we take a more comprehensive approach 
and use the full shape of the marked CF to predict the cosmological parameters.
Figure \ref{fig_cov} shows the {\it correlation coefficients } 
of $\hat W(s)$ and $\hat W_{\Delta s}(\mu)$
\footnote{The covariance matrix of different $\alpha$s have very different magnitude, 
so we plot the correlation coefficients.}.
They are estimated using the 150 COLA simulations.
Among all MCFs, the $\alpha=-1$ case has weakest correlation with the others.
The negative correlations are the consequence of the normalization.


We choose the $\Omega_m=0.3071,\ w=-1$ cosmology as the {\it fiducial geometrical background},
and define a statistical function to 
distinguish the other backgrounds from it,
\begin{equation}
 \chi^2 = ({\bf p_{\rm fiducial}-p_{\rm target}}) \cdot {\bf Cov}^{-1} \cdot ({\bf p_{\rm fiducial}-p_{\rm target}}),
\end{equation}
where $\bf p$ denotes $\hat W(s)$ or $\hat W_{\Delta s}(\mu)$.
Considering that the number of mocks 
is not too many compared with the binning number of the $W$s,
we use the formula suggested by \citet{Hartlap:2006kj} to correct 
the bias in the estimated covariance matrix.


The MCFs of the halo catalogues embedded in the backgrounds of the incorrect cosmologies are obtained using
the following coordinate transforms (see \cite{LI18} for details),
\begin{equation}\label{eq_smu_trans1}
 s_{\rm target} = s_{\rm fiducial} 
 \sqrt{\alpha_{\parallel}^2\mu_{\rm fiducial}^2 + \alpha_{\bot}^2(1-\mu_{\rm \bot}^2)},
\end{equation}
\begin{equation}\label{eq_smu_trans2}
 \mu_{\rm target} = \mu_{\rm fiducial} \frac{\alpha_\parallel}
 {\sqrt{\alpha_{\parallel}^2\mu_{\rm fiducial}^2 + \alpha_{\bot}^2(1-\mu_{\rm \bot}^2)}}.
\end{equation}
This is much more efficient compared with converting the samples into the different backgrounds and re-measuring the MCFs.
A caveat is that equations \ref{eq_smu_trans1} and  \ref{eq_smu_trans2} do not capture 
{\it the change in the values of the weights}.
That definitely happens, since the Alcock-Paczynski effect non-uniformly distorts the geometry,
so the set of $n_{\rm NB}$ nearest neighbors can differ from one cosmology to next.
In Appendix C we check this caveat and show that if we neglect this issue it introduces only minor effect, and 
thus equations \ref{eq_smu_trans1} and  \ref{eq_smu_trans2} are deemed precise enough for this proof-of-concept study.

Equations \ref{eq_smu_trans1} and \ref{eq_smu_trans2} only consider the {\it background} 
information of the cosmologies.
A more comprehensive analysis should involve the information of the structure growth,
but that would require many more numerical simulations.

\subsection{Constraints from $\hat W(s)$}

The conditional constraints on $\Omega_m$ and $w$ (fixing one of them as the fiducial and constrain the other one) 
using the full shape of $\hat W_{\Delta s}(\mu)$ 
are presented in the lower panel of Figure \ref{fig_chisq_1d}.
In the plots, we use the clustering range of $s\in(5,50)\ h^{-1}\rm Mpc$, divided into 15 bins.
Including the results on larger scales does not further enhance the power of constraints.

We find the the $\alpha=0$ results leads to the tightest constraints
among all cases considered.
Also, combining different MCFs can improve the constraint.
Taking the $w=-0.4$ cosmology as an example.
Compared with the fiducial cosmology, it is disfavored by 
$\chi^2=7.3/5.8/5.5/1.4$
when using the $\alpha=0/0.5/-0.5/1$ MCF,
so the $\alpha=0.5/-0.5/1$ result is
20\%/24\%/81\% worse than the $\alpha=0$ result, respectively.
Combining the $\alpha=0$ and $\alpha=1$ MCFs,
we get a 17\% improvement compared with only using the $\alpha=0$ MCF.
If we combine the  $\alpha=0/0.5/1$ MCFs together,
the $\chi^2$ is then enlarged to 15,
a $\approx$100\% improvement compared with only using the $\alpha=0$ MCF.

The $\chi^2$ of the $\alpha=0,\ 0.5,\ 1$ combination 
is very close to the summation of the $\chi^2$s using the three MCFs separately.
This means that, 
the cosmological information carried by the three MCFs
is not strongly overlapping from each other.
This is essentially important for the MCF statistics,
meaning that we can significantly improve the cosmological constraints by combining different MCFs.



\subsection{Constraints from $\hat W_{\Delta s}(\mu)$}

\begin{figure}
      \centering
      \includegraphics[width=8cm]{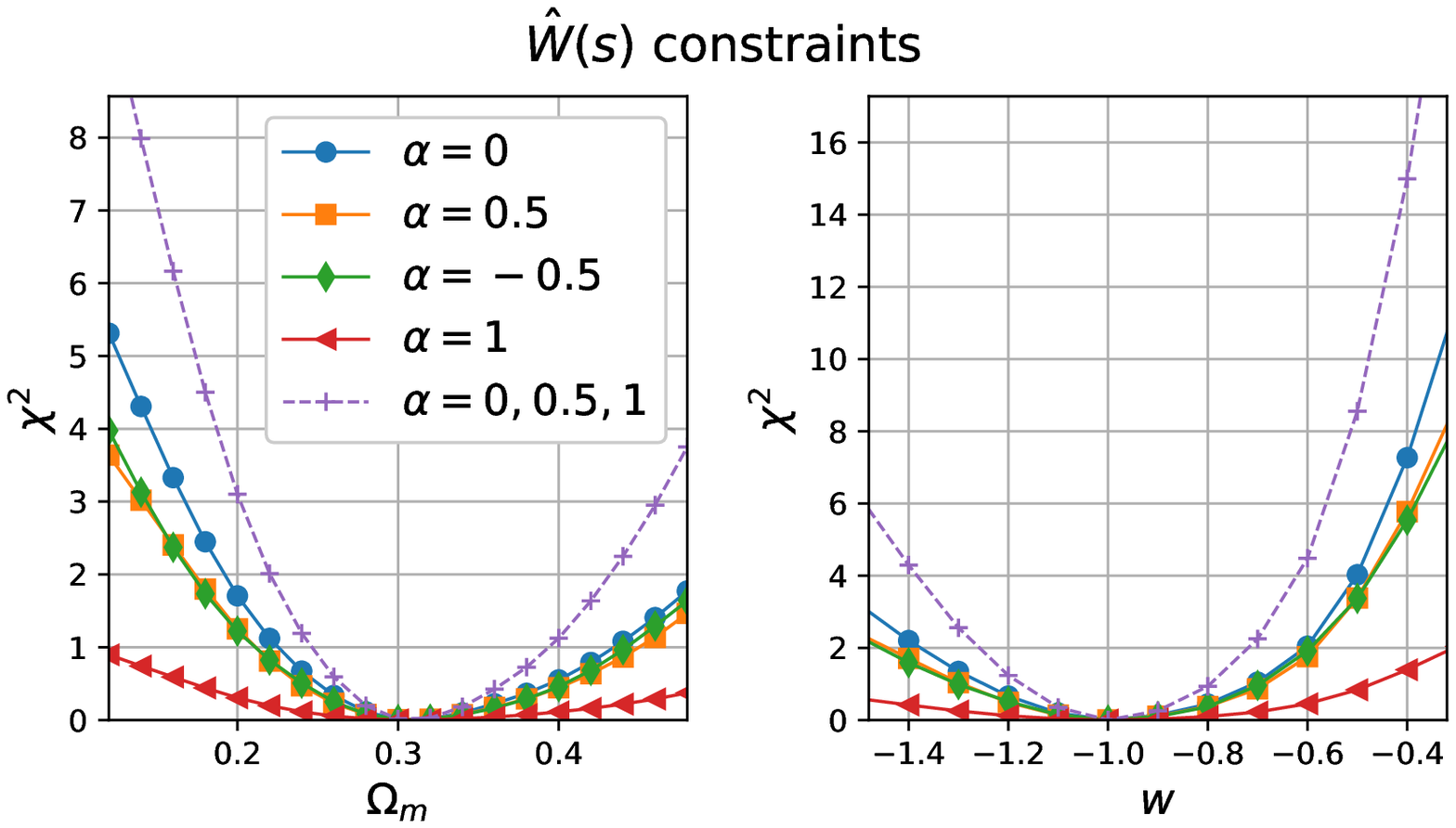}
      \includegraphics[width=8cm]{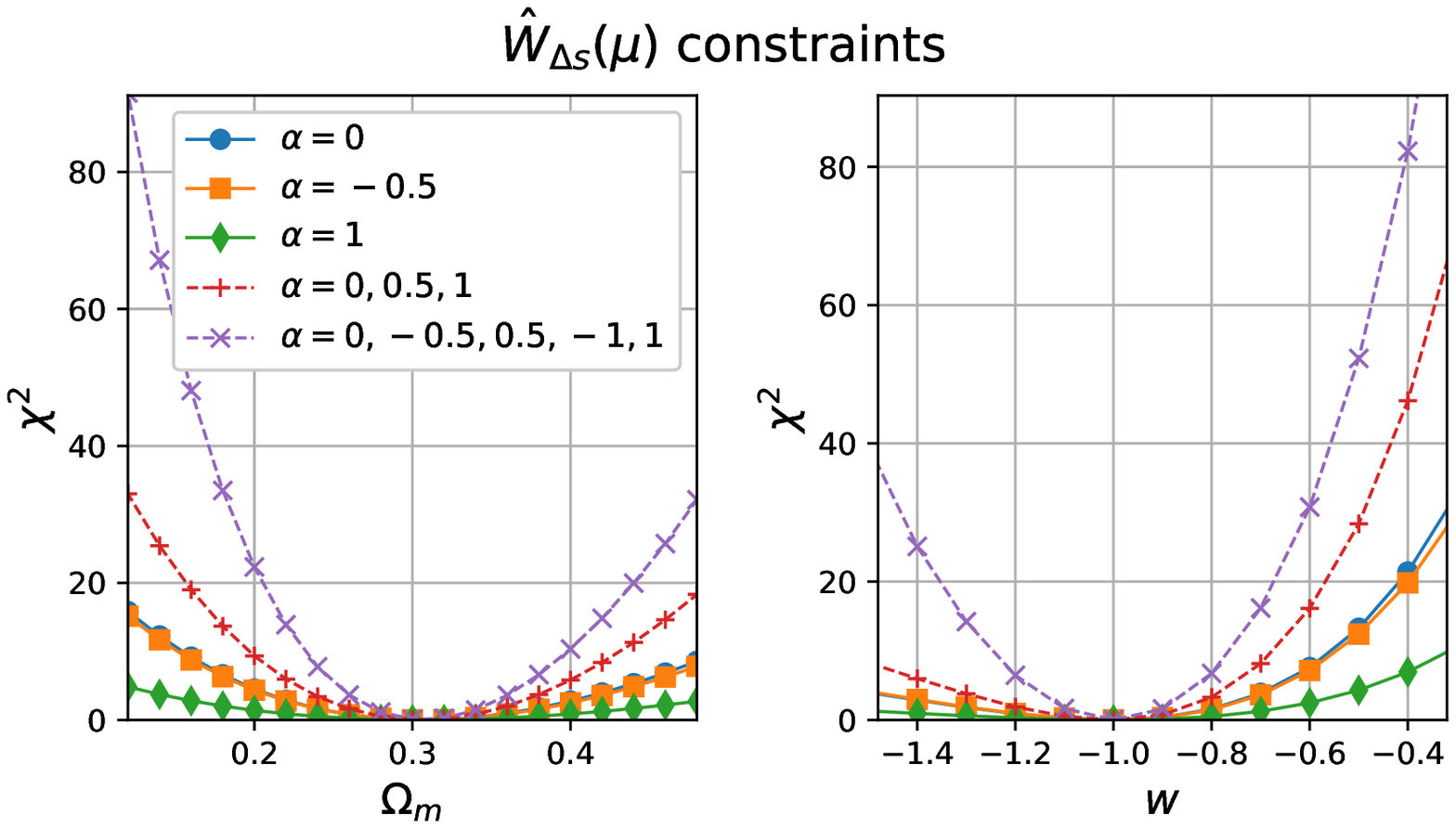}
      \caption{Conditional constraints on $\Omega_m$ and $w$, 
       derived using $W_{\Delta s}(\mu)$ and $\hat W_{\Delta s}(\mu)$.
       Estimation is based on the $\bar n=10^{-3}\ (h^{-1}\rm Mpc)^3$ halo samples of
       the 150 COLA samples having a boxsize of $(512 h^{-1}\rm Mpc)^3$.
       Among all MCFs, the $\alpha=0$ MCF has the largest statistical power.
       We can largely improve the statistical power by combining different MCFs.
      }      \label{fig_chisq_1d}
\end{figure}

The conditional constraints on $\Omega_m$ and $w$ using the full shape of $\hat W_{\Delta s}(\mu)$
are presented in the lower panel of Figure \ref{fig_chisq_1d},
where the integration range of $s$ is taken as $(6,40)\ h^{-1}\rm Mpc$,
and we use the shape of $\hat W_{\Delta s}(\mu)$ in the range of 
$\mu\in(0,0.97)$, divided into 12 bins.
Remarkably, the constraints derived using the $\hat W_{\Delta s}(\mu)$s 
are much more powerful than those derived using the $\hat W(s)$s.
 
Similarly, we find $\alpha=0$ achieves the best performance,
and the results can be improved by combining different MCFs.
In Table \ref{tab1}, we  list the $\chi^2$s of the $w=-0.4$ cosmology using different $\alpha$s or their combinations.
Compared with using $\alpha=0$ MCF, using $\alpha=0,0.5,1$ and $\alpha=0,-1,-0.5,0.5,1$
can improve the $\chi^2$ by 116\% and 285\%, respectively.

\begin{figure}
      \centering
      \includegraphics[width=8cm]{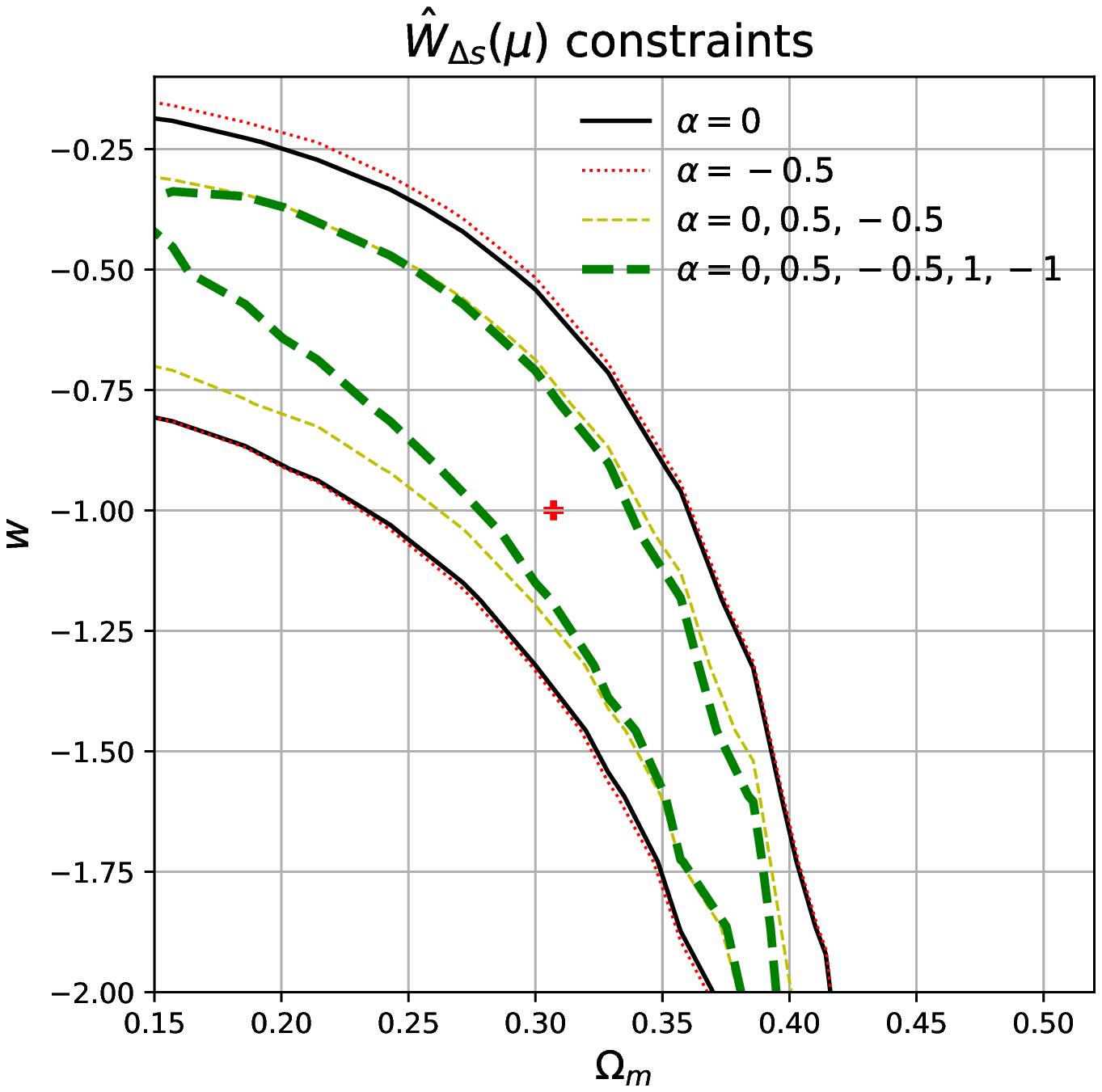}
      \caption{\label{fig_chisq_contour}
      68.3\% CL constraints on $\Omega_m$ and $w$, derived using $\hat W_{\Delta s}(\mu)$. 
      Different MCFs have the same direction of degeneracy.
      Compared with only using one kind of MCF,
      the two combinations can significantly reduce the constrained area. }
\end{figure}

Figure \ref{fig_chisq_contour} shows the constraints in the 2-d $\Omega_m$-$w$ parameter space.
The directions of degeneracy using different $\alpha$s are identical to each other,
so combining the different MCFs does not help in breaking the degeneracy.
But by doing this we do manage to shrink the contour size.
Very roughly, compared with the $\alpha=0$ MCF,
the $\alpha=0,0.5,1$ and $\alpha=0,-1,-0.5,0.5,1$ combinations can
improve the constraints on the parameters by $\approx30\%$ and 50\%, respectively.

\begin{table*}
\begin{center}
\caption{
\textrm{\ \ \ \ $\chi^2$s of the $w=-0.4$ cosmology, derived using $\hat W_{\Delta s}(\mu)$ with different $\alpha$ or their combinations }}
\label{tab1}
\begin{tabular}{c|ccccccccccc}
  \hline
   $\alpha$     &  0 & -1 & -0.5 & 0.5 & 1 & 0,-1 & 0,-0.5 & 0,0.5 & 0,1 
   & 0,0.5,1 & 0,-1,-0.5,0.5,1 \\
   \hline
   $\chi^2$ &  21.4 & 8.1 & 19.8 & 14.9 & 6.9 & 29.2 & 45.0 & 40.0 & 30.8 & 46.2 & 82.3   \\
     \hline
    $\frac{\chi^2}{\chi^2_{\alpha=0}}-1$ &  0\%  & -62\% & -7\% & -31\% & -68\% & 36\% & 111\% & 87\% & 44\% & 116\% & 285\%    \\
  \hline
\end{tabular}
\end{center}
\end{table*}

\section{Conclusion}\label{conclusion}

We performed a detailed analysis on the MCFs for which the objects are weighted by $\rho^\alpha$.
In this analysis, we considered five different MCFs, i.e. $\alpha=-1,\ -0.5,\ 0,\ 0.5$ and 1,
and characterize their scale and angular dependence by using $W(s)$ and $W_{\Delta s}(\mu)$.
When studying the scale dependence of the MCFs, i.e. the $W(s)$s, 
we find the different MCFs have very different amplitudes and scale-dependence.
Especially, we found distinctive peaks and valleys in some $\alpha\neq 0$ MCFs,
on scales around $\approx5$ and $15$ $h^{-1}$ Mpc, 
depending on the smoothing scale that we adopted to estimate the density.
Their origin and properties are studied in detail.
One particular point of interest is that the locations of these features are rather invariant with redshift.

In studying the possibilities of using the MCFs in cosmological analysis,
we find the locations of the peaks or valleys are rather insensitive to the background geometry.
Thus, it is unlikely that they can be utilized as ``standard rulers'' to probe the geometry.
However, their locations are affected by the value of $\sigma_8$ and the galaxy bias,
so they could be useful for the determination of these parameters.

Finally, we studied the power of the different MCFs in distinguishing the different cosmologies,
by using the full shape of the $\hat W(s)$s and the $\hat W_{\Delta s}(\mu)$s.
We find they have similar direction of degeneracy in constrained $\Omega_m$ and $w$,
while the $\alpha=0$ MCF, corresponding to the standard CF,
has the strongest power in distinguishing the background of the different cosmologies.
Also, the constraint can be further improved by combining the different MCFs together.
In particular, compared with the $\alpha=0$ $W_{\Delta s}(\mu)$,
the $\alpha=0,\ 0.5,\ 1$ and $\alpha=0,\ -1,\ -0.5,\ 0.5,\ 1$ combinations achieve 
$\approx30\%$ and 50\% improvement in reducing the constrained area, respectively.

The reason why MCF can improve the constrain is easy to understand.
The dense and under-dense regions have very different clustering patterns and RSDs features.
The many MCFs provide different weighting schemes of the clustering information 
according to their local density.
By using them together, we can separate the regions with different patterns 
and extracting more clustering information.

While previous works regarding the MCF mainly focus on modified gravity theories\citep{2018Aguayo,2018Armijo},
our work suggests that they could be useful for probing any parameter
that is related with the expansion and structure growth history.
By using the MCFs, we can enlarge the obtained information by 3-4 more times.
MCF are also computationally efficient  compared with the high order statistics, like the 3pCF. 

\cite{PMS2020} used perturbation theory to study the marked power spectrum using perturbation theory,
and found that the mark introduces a significant coupling between small-scale non-Gaussianities and large scale clustering.
This explains why using this statistics we can get additional information,
and provides further support to the findings of this work.

We find that the statistical quantity $\hat W_{\Delta s}(\mu)$ is more powerful than 
the $\hat W(s)$ in constraining the cosmological parameters.
It may be possible to use  $\hat W_{\Delta s}(\mu)$s instead of just $\hat \xi_{\Delta s}(\mu)$ in the tomographic AP method
to improve the performance. However we leave this issue for future works.

While in our analysis the $\alpha=0$ MCF has the strongest power in distinguishing the background of the different cosmologies,
the authors of \cite{2020arXiv200111024M} found that, 
a marked power spectrum can better constrain cosmological parameters than the power spectrum itself.
This difference may be due to two reasons.
1) By using tens of thousands of simulations, \cite{2020arXiv200111024M} built an emulator to 
capture both the expansion history and structure formation of the Universe.
In contrast, we just ``re-observe'' one simulation using different backgrounds
to study effect of the expansion history.
Very possibly, the sensitivity of the MCFs to the structure formation is more important than its dependence to the expansion history,
but we do not have it quantified in this simple treatment.
We need to conduct a more comprehensive study in future analysis.
2) While \cite{2020arXiv200111024M} used the power spectrum as the statistical discriminator, 
we used $W_{\Delta s}(\mu)$, the dependence of clustering strength on the direction. 
The two statistical quantities are physically quite different, 
and it is reasonable that the results derived using them are also different.

There are still many issues regarding MCF that are important but that we chose not to address in the present work.
Although we have shown that MCFs encode a lot of information,
we did not detail specific methods to extract them.
In particular, we did not check whether the MCFs is useful for improve the measuring of the BAO peaks.
In studying the different weighting schemes we only explore the restricted range of $-1\leq\alpha\leq1$.
Finally, we only considered the halo number density as the weight, while there are possibilities to use 
features computed directly on the connectivity graph of the halo distribution. 
Those graph features are related to topological characteristics of the cosmic web \citep{Suarez-Perez:2020}, features that 
are in turn naturally correlated to $\sigma_8$ and the halo-galaxy bias.

\section*{Appendix A. Usability of Peaks in the $\beta$-skeleton statistics}\label{appa}

The $\beta$-skeleton is a novel statistical tool proposed in \cite{Fang2019} 
to study the cosmic web.
In this statistic, the ``spikes'' produced by the FOG 
leads to a peak in the histogram of connections near $2.5{h^{-1}\rm Mpc}$,
which is rather robust to the redshift.
The origin and the properties of this peak is very close to 
to the peak we found in the $\alpha=1,\ 0.5,\ -1$ $W(s)$s.

\begin{figure}[htbp]
 \centering
      \includegraphics[width=9cm]{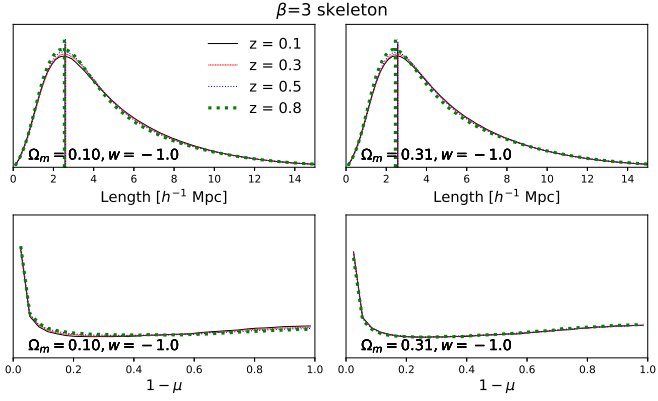}
      \caption{      
      Histograms of lengths and directions of connections in the $\beta=3$ cosmic web.
      Similar to the $W(s)$, there are peaks produced by the FOG on the small scales,
      which are robust against the redshift, but unlikely to be used in cosmological studies.
      The full shape of the histograms show some cosmological dependence.
      }      \label{fig_beta_hists}
\end{figure}

We find that, the peak in the $\beta$-skeleton statistics also cannot be used to probe the cosmic expansion history.
Figure \ref{fig_beta_hists} shows the 
distribution of the lengths and directions of the connections in the $\beta=3$ web, 
measured in three different backgrounds.
It is clear that the locations of the peaks are 
rather insensitive to the background geometry.

The full shape of the histograms of the lengths and directions 
show some cosmological dependence.
We will not go deep and discuss its usability in details.

In other work we have found that the entropy and complexity of the $\beta$-skeleton graph actually correlates
with $\sigma_8$ \citep{Torres-Guarin:2020}, suggesting that the graph is more sensitive to the global tracer topology than to the 
more geometrical influence of $\Omega_m$ and $w$.

\section*{Appendix B. $W_{\Delta s}(\mu)$ in the five COLA simulations}
\label{appb}

In section \ref{sec_five_cola_sim} we only discussed the 
$W(s)$ measured in simulations with different  $\sigma_8$.
Here we present the results of $\hat W_{\Delta s}(\mu)$s. 

\begin{figure}[htbp]
      \centering
      \includegraphics[width=8cm]{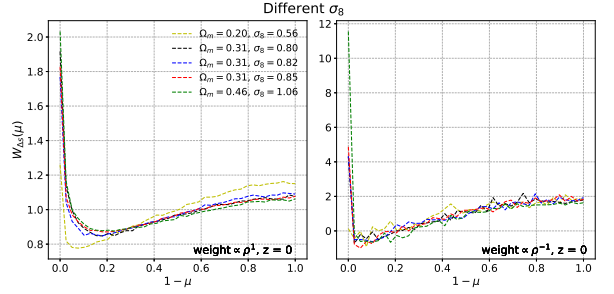}
      \caption{
      The $W_{\Delta s}(\mu)$s measured in the five sets of COLA simulations with different values of $\sigma_8$.  
      The leftmost part of the curve is dominated by the FOG effect.
      There the amplitude is significantly enhanced if using a large $\sigma_8$.
      }      \label{fig_ximu_diffcosmo}
\end{figure}

As shown in Figure \ref{fig_ximu}, 
the $1-\mu\lesssim0.1$ part, where the FOG should dominate, 
has a strong dependence on the value of $\sigma_8$.
A larger $\sigma_8$ results in a stronger FOG effect, and thus a sharper peak.

The $1-\mu\gtrsim0.1$ part, dominated by the Kaiser effect,
seems to have a similar shape with different $\sigma_8$.

\section*{Appendix C. Accuracy of the Approximately Estimated MCFs}

\begin{figure}
      \centering
      \includegraphics[width=8cm]{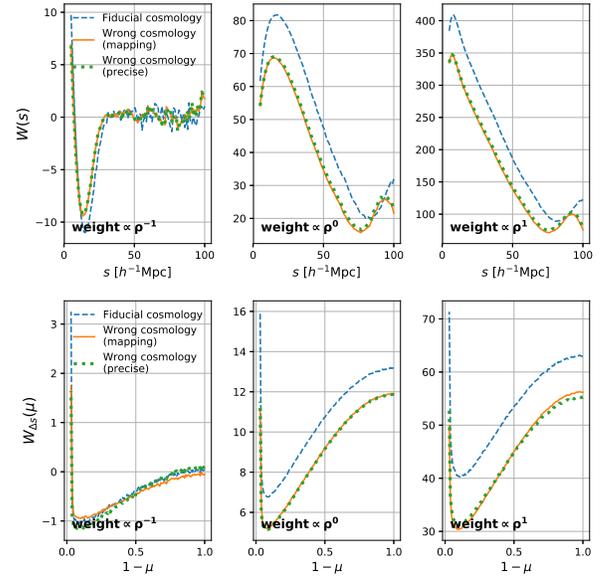}
      \caption{
      $W(s)$ and $W_{\Delta s}(\mu)$ of the BigMD $z=0.5$ halos, 
      measured in the backgrounds of the fiducial cosmology and a wrong one.
      Using  equations \ref{eq_smu_trans1} and  \ref{eq_smu_trans2}, we can estimate the wrong cosmology results in a fast speed
      while still maintaining an enough accuracy.
      }      \label{fig_wrong_cosmos}
\end{figure}

Figure \ref{fig_wrong_cosmos} shows $W(s)$, $W_{\Delta s}(\mu)$ in the fiducial cosmology 
$(\Omega_m,w)=(0.3071,-1)$,
and in the background of a wrong cosmology $(\Omega_m,w)=(0.5,-1)$.
The results in the wrong background are computed in two ways,
the precise measurement obtained by constructing the sample in the wrong background and then re-measuring the MCFs,
and also the approximate results inferred using equations \ref{eq_smu_trans1} and  \ref{eq_smu_trans2},
Inspected by eye, we find the approximate results are very close to their precise correspondance.

Since equations \ref{eq_smu_trans1} and  \ref{eq_smu_trans2} do not capture 
the change of the weights in the different backgrounds,
it is important to check its influence.
Here we showed that it is minor compared with the cosmological effect.



\section*{acknowledgments}


We thank Kwan-Chuen Chan and Xin Wang for helpful discussions. 
XDL acknowledges the supported from NSFC grant (No. 11803094), 
the Science and Technology Program of Guangzhou, China (No. 202002030360).
CGS acknowledges financial support from the National Research Foundation of Korea
(NRF; \#2020R1I1A1A01073494).
J.E. F-R acknowledges support from COLCIENCIAS Contract No. 287-2016,
Project 1204-712-50459.  
We acknowledge the use of 
the {\it Kunlun} cluster, a supercomputer owned by the School of Physics and Astronomy, Sun Yat-Sen University.

The CosmoSim database used in this paper is a service by the
Leibniz-Institute for Astrophysics Potsdam (AIP).
The MultiDark database was developed in cooperation with
the Spanish MultiDark Consolider Project CSD2009-00064.

\bibliographystyle{aasjournal}
\bibliography{cites}


\end{document}